\documentclass{ws-procs961x669}            
\begin{document}
\title{Cyclic Cosmology and Holographic Entanglement Entropy}

\author{P.H. Frampton$^*$}

\address{Dipartimento di Matematica e Fisica "Ennio De Giorgi",\\ 
Universita del Salento and INFN-Lecce,\\ Via Arnesano, 73100 Lecce, Italy.\\
$^*$E-mail: paul.h.frampton@gmail.com}

\begin{abstract}

\end{abstract}


\bodymatter

\section{Introduction}
\noindent
Cyclic cosmology with an infinite number of cycles each of which includes expansion, 
turnaround, contraction and bounce, is an appealing theory, and one of the 
most recent attempts at a solution was in 2007 by Baum and me
\cite{BaumFrampton}. As 
emphasized first by Tolman, entropy and the second law of thermodynamics 
provide a significant challenge to implementing a cyclic theory\cite{Tolman}.

\bigskip

\noindent
The approach \cite{BaumFrampton} was based on the so-called Come Back Empty (CBE) principle which chooses
the turnaround as the point in the cycles where entropy is jettisoned. A successful universe must be selected with zero entropy at the turnaround to begin an adiabatic contraction to the bounce while empty of all matter including black holes.
In this article we provide a better understanding of CBE based on the holographic principle. In particular, the notion of entanglement entropy and the related question of spacetime connectivity can put the CBE model on a quantum theoretical basis.

\bigskip

\noindent
Although the holographic principle was well-known at that time, its relationship to cyclic cosmology was unclear. Even how it could help reconcile general relativity with quantum mechanics was less developed. In the AdS/CFT correspondence, it has been argued that the dynamics of spacetime on the AdS side, including the connectivity of spacetime, is related to quantum entanglement of disconnected classi-cal theories on the CFT side. This idea was introduced by Van Raamsdonk in an essay submitted to the Gravity Research Foundation\cite{VanRaamsdonck}. Considerations of the entropy of the universe have been discussed also by E. Verlinde\cite{ErikVerlinde}.

\bigskip

\noindent
Before discussing the impact on cyclic cosmology which is our present subject, let us briefly discuss how this line of reasoning changes the marriage of general relativity with quantum mechanics. For several decades it had been tacitly assumed that there exists a quantum gravity theory whose classical limit is general relativity. Now it appears that whether or not such a theory exists may not be the best question to ask. Classical general relativity which describes the dynamics of spacetime may instead arise from consistency conditions on the quantum entanglement of conformal field theories which are at the holographic boundary.

\bigskip

\noindent
Subsequent work has been mainly on time-independent spacetimes. However, we shall show that for the the time-dependent CBE model of cyclic cosmology use of holographic quantum entanglement entropy makes possible a better understanding of entropy at the turnaround.
First we briefly review the CBE model.
\bigskip

\noindent
The best way to discuss the CBE model is by using a language where the spacetime occupied by the universe is divided into two subregions denoted by introverse and extroverse respectively which we must define carefully because it will be the entanglement entropy between these two subregions which will play a central r\^ole in the cyclicity.
The present visible universe at $t = t_0$ is the present introverse and its comoving radius is
$R_{int} (t_0) = 44Gly$.
A more useful time to consider than the present $t_0 = 13.8Gy$ is the time $t = t_{DE}$ 
when the dark energy began to dominate over the matter component. This occurred at 
$t_{DE} = 9.8Gy$ at which time the extroverse is identified with the introverse.

\bigskip

\noindent
After $t=t_{DE}$, the extroverse expands exponentially and become larger, eventually
very much larger, than the introverse. 

\bigskip

\noindent
At $t=t_{DE}$ we have

\begin{equation}
R_{ext}(t_{DE}) = R_{int}(t_{DE}) = 39Gly
\label{tDE}
\end{equation}

\noindent
at a time when the scale factor, normalized to $a(t_0)=1$, was $a(t_{DE})=0.75$.

\bigskip

\noindent
The extroverse expands exponentially for $t_{DE} < t < t_T$, where $t_T$ is the turnaround
time calculated to be $t_T = 1.3$Ty. The extroverse comoving radius during this
expansion is given by

\begin{equation}
R_{ext}(t) = \left( 39 Gly \right) {\rm exp} \left[ \frac{t-9.8Gy}{13.8Gy} \right]
\label{Rext}
\end{equation}

\noindent
Two interesting values at $t=t_0$ and $t=t_T$ respectively are

\begin{equation}
R_{ext}(t_0) = 52 Gly ~~~~~ {\rm and} ~~~~~ R_{ext}(t_T) = 1.5 \times 10^{42} Gly
\label{Rext}
\end{equation}

\bigskip

\noindent
The introverse expands more gradually for $t_{DE} < t < t_T$, being limited by the speed
of light in its definition. The introverse comoving radius for the expansion era is given in terms
of $a(t) = {\rm exp} [ (t - 13.8 Gy)/(13.8 Gy)]$ by

\begin{eqnarray}
R_{int}(t) & = & 39 Gly + c \int_{t_{DE}}^{t} dt a(t)^{-1}   \nonumber  \\
&= & 39 Gly + \nonumber \\
& &\left(13.8 Gly\right) \left[1-{\rm exp}\{-(t-9.8Gy)/(13.8Gy)\}\right] \nonumber \\
&=&44Gly+\nonumber \\
& &\left(13.8 Gly\right) \left[1-{\rm exp}\{-(t -13.8Gy)/(13.8Gy)\}\right]  
\label{Rinverse}
\end{eqnarray}
   
\noindent
From Eq.(\ref{Rinverse}) the values of $R_{inv}(t)$ at $t=t_0$ and $t=t_T$ are

\begin{equation}
R_{int}(t_0) = 44 Gly ~~~~~ {\rm and} ~~~~~ R_{inv}(t_T) = 58 Gly
\label{Rint}
\end{equation} 

\bigskip

\noindent
The behaviour of $R_{ext}(t)$ and $R_{inv}(t)$ calculated in (\ref{Rext}) and (\ref{Rint})
respectively assumes that the dark energy is accurately described by a cosmological constant.  
The results for $R_{ext}(t_0)$ and $R_{inv}(t_0)$ show that the present extroverse is 
already $60\%$ larger in volume
than the introverse implying that hundreds of billions of galaxies have exited the introverse
since the onset of exponential expansion.

\bigskip

\noindent
We have previously used a precise value for the turnaround time
$t_T = 1.3Ty$ without further comment so we provide a few more words.
For infinite cyclicity, it is necessary to impose a matching condition on the 
contraction scale factor 
with the scale factor of the previous expansion. Since the contraction 
has a radiation-dominated behaviour $\propto \sqrt{t}$ the matching 
is conveniently made at the onset of matter domination during expansion 
when $t = t_m = 47ky$. This leads to $t_T = 1.3 Ty$ for the turnarounds time. 
We note that this could not be derived in the BF model\cite{BaumFrampton}
because there the equation of state for the dark energy was assumed to be 
$\omega < -1$ but cyclicity is now known to require $\omega = - 1$ precisely.

\bigskip

\noindent
We shall improve the original version of Comes Back Empty (CBE) with a new understanding based on quantum entanglement. Although quantum fluctuations as precursors of cosmic structure formation are well-known, to our knowledge this is the first application of quantum computation and quantum information, in par- ticular quantum entanglement, to theoretical cosmology. To see that it is an improvement, let us first describe the original CBE model.

\bigskip

\noindent
In the original derivation which used phantom dark energy with $\omega < -1$, a modified Friedmann equation in the form

\begin{equation}
\left( \frac{\dot{a}}{a} \right)^2 = \frac{8 \pi G}{3}
\left[ \rho - \frac{\rho^2}{\rho_{cr}} \right]
\label{FriedmannString}
\end{equation}

\noindent
was used where $\rho_{cr}$ is a constant. Such a modified Friedmann equation
was derived from string theory. 
This led to a bounce when $\rho = \rho_{cr}$ at a time $< 10^{-27}s$ before the Big Rip. 
The scale factor deflated dramatically to $\hat{a}(t_T) = f a(t_T)$ where $f < 10^{-28}$.
All bound states become unbound and soon afterwards become causally disconnected. The critical density is $\rho_{cr} = \eta\rho_{H_2O}$ with $\eta$ between
$10^{31}$ and $10^{87}$, safely below the Planck value $\eta = 10^{104}$. A given patch is chosen so that it is empty of all matter including black holes, and contains only dark energy. This causal patch contracts adiabatically to a bounce with zero entropy.

\bigskip

\noindent
In the present derivation, the dark energy is described by a cosmological constant with 
$\omega = -1$ so that we must use the unmodified Friedmann equation

\begin{equation}
\left( \frac{\dot{a}}{a} \right)^2 = \frac{8 \pi G}{3}
\rho
\label{Friedmann}
\end{equation}

\noindent
at almost all times except times extremely close to the turnaround or the bounce. 

\bigskip

\noindent
Now we turn to the r\^{o}le of quantum entanglement. This begins from the holographic principle and its realization in the AdS/CFT duality. One previous use of quantum mechanics in cosmology has been the idea that quantum fluctuations of the inflaton field during inflation are amplified to seed large scale structure 
formation. Here we suggest that quantum entanglement and its relationship with spacetime connectivity play an comparably dramatic r\^{o}le by elucidating the CBE cyclic cosmology model at turnaround.

\bigskip

\noindent
In the AdS/CFT correspondence, if we consider two 
non-interacting identical copies $CFT_A$ and $CFT_B$, a state of the
system can in general be written
\begin{equation}
|\Psi> = |\Psi>_A \otimes |\Psi>_B.
\label{product}
\end{equation}
The CFTs are each dual to separate asymptotically AdS spacetimes
so that the direct product (\ref{product}) is dual to two spacetimes which
are disconnected.

\bigskip

\noindent
As a quantum state we are free to consider a superposition of states.
Let the energy eigenstates be $|E_k>$ for one CFT and consider the 
quantum superposition

\begin{equation}
|\Psi (\beta) > = \Sigma_k e^{-\frac{\beta}{2}E_k} |E_k>\otimes|E_k>
\label{superposition}
\end{equation}

\noindent
which can be shown, in general, to be dual to a single connected spacetime.
Starting with $|\Psi>$ we deduce that the denstiy matrix for $CFT_B$ is

\begin{equation}
Tr \left( |\Psi> <\Psi| \right) = \Sigma_k e^{-\beta E_k} |E_k><E_k| = \rho_T
\label{thermal}
\end{equation}

\noindent
which is a thermal density matrix. so the quantum superposition of two
disconnected spacetimes is identified with a classical connected spacetime.
This leads to the fascinating idea that 
{\it classical spacetime connectivity arises by quantum entangling the degrees of freedom in two components}

\noindent
which is a central idea in what follows.

\bigskip

\noindent
Proceeding in the opposite direction, let us consider one CFT on a sphere and divide the sphere
initially into two hemispheres A and B with Hilbert space ${\cal H}_A\otimes{\cal H}_B$.
The entanglement entropy is the von Neumann entropy
\begin{equation}
S(A) = - Tr \left( \rho_A {\rm log} \rho_A \right)
\label{vonNeumann}
\end{equation}

\noindent
with

\begin{equation}
\rho_A = Tr_B \left( |\Psi><\Psi| \right).
\label{rhoA}
\end{equation}

\noindent
 As $S(A)$ decreases, the area of the minimum surface separating A and B decreases
 according to the Ryu-Takayanagi prescription which greatly generalises
 the original Bekenstein-Hawking area formula for the entropy of a black hole.
 The sphere becomes
 a dumbbell shape. In the limit $S(A) \rightarrow 0$, the spacetimes A and B become 
 classically disconnected.

\bigskip

\noindent
Let us briefly consider entanglement of two-state quantum mechanical systems,
or quantum bits usually shortened to qubits. One qubit can be in the general state

\begin{equation}
\Psi = \alpha_{\uparrow} |\uparrow> + \alpha_{\downarrow}|\downarrow>
\label{onequbit}
\end{equation}

\noindent
with probabilities $|\alpha_{\uparrow}|^2$ and $|\alpha_{\downarrow}|^2$ of measuring $|\uparrow>$ and
$|\downarrow>$ respectively.

\bigskip

\noindent
For two qubits the most general state is

\begin{equation}
|\Psi> = \alpha_{\uparrow\uparrow}|\uparrow\uparrow> + \alpha_{\uparrow\downarrow}|\uparrow\downarrow> +
\alpha_{\downarrow\uparrow}|\downarrow\uparrow> +
\alpha_{\downarrow\downarrow}|\downarrow\downarrow> 
\label{twoqubits}
\end{equation}

\noindent
with $\Sigma |\alpha_{ij}|^2 = 1$.

\bigskip

\noindent
If we measure the first qubit to be $|\uparrow>$, then the normalized post-measurement state
is

\begin{equation}
|\Psi>^{'} = \left[ \frac{\alpha_{\uparrow\uparrow} |\uparrow\uparrow> + 
\alpha_{\uparrow\downarrow} |\uparrow\downarrow>}
{\sqrt{|\alpha_{\uparrow\uparrow}|^2 + |\alpha_{\uparrow\downarrow}|^2}} \right]
\label{postmeasure}
\end{equation}

\bigskip

\noindent
The phenomenon of entanglement is clearest in an EPR state

\begin{equation}
|\Psi> = \frac{1}{\sqrt{2}} \left[ |\uparrow\uparrow> + |\downarrow\downarrow> \right]
\label{EPR}
\end{equation}

\noindent
named for a paper which tried unsuccessfully to prove that
quantum mechanics is incomplete. Eq.(\ref{EPR}) is sometimes alternatively called a Bell
state for the physicist who derived inequalities based on local realism
which are violated by quantum mechanics.
The correlations which occur in quantum mechanics are stronger than
could ever exist between classical systems. The point 
about the EPR/Bell state Eq.(\ref{EPR}) is that measuring the first qubit as $|\uparrow>$
implies that measurement of the second qubit must with 100\% certainty result 
in $|\uparrow>$, a correlation which never happens in classical mechanics.

\bigskip

\noindent
After these digressions, we return to our central point of the turnaround in cyclic cosmology.
The CBE model with its empty contraction era aims to jettison 
the accrued entropy at the turnaround so that the contracting
universe has zero entropy. One reason for this is that the presence of matter including black holes
will disallow contraction to the bounce because black holes will merge and enlarge and 
in the presence of any matter going in reverse
through phase transitions will be entropically impossible. Thus the idea was that the causal
patch at turnaround which is selected from a very large number of candidates, as suggested
by the comparison between the two turnaround radii in Eqs.(\ref{Rext})
and (\ref{Rint}), will almost always be empty of matter. The tiny fraction which do contain matter
will represent failed universes which will experience a premature bounce rather than a
successful contraction.

\bigskip

\noindent
Mathematicians or string theorists may provide details of a non-singular turnaround
but until then we remain temporarily satisfied with the phenomenological Eq.(\ref{remodified})
with a view first to discover a scenario which {\it can} be infinitely cyclic.

\bigskip

\noindent
As the extroverse spacetime is stretched by the extreme amount in Eq.(\ref{Rext})
it will become disconnected into causal patches and this disconnection
corresponds precisely to making $S(int)$ vanish where $S(int)$ is the
entanglement entropy between introverse and extroverse.
Because the extroverse becomes a classically disconnected spacetime, we may
study the contracting introverse without any further concern for the extroverse
whose entropy is thereby jettisoned. 

\bigskip

\noindent
Tolman's work had so much impact that almost
no work on cyclic cosmology was pursued thereafter. The most important subsequent step was the discovery
of dark energy in 1998 which provides the clear division into introverse and extroverse as discussed
in the previous section. Tolman was necessarily considering the entropy
for the whole universe and applied the second law of thermodynamics to it, to arrive at
his impossibility theorem.

\bigskip

\noindent
Tolman was certainly aware of quantum mechanics but it surely never entered his head
that quantum mechanics is relevant to cosmology. Such an idea only emerged long after
Tolman died in 1948, in the 1980s when quantum fluctuations were proposed to seed
structure formation. Now we
learn that, at the other end of the expansion era, quantum mechanics may play
a r\^ole also in the spacetime disconnection of the introverse, as well as 
in choosing the quantum entanglement entropy of the introverse to be
the quantity which is infinitely cyclic.

\bigskip

\noindent
As for experimental tests of CBE cyclic cosmology, regrettably only one is immediately
apparent which is the
prediction that the equation of state of dark energy is a constant
$\omega = -1$ very precisely to many decimal places. If a measurable departure
from this prediction were found experimentally, it would refute the CBE 
cyclic model. Hopefully with more work on the model, other testable
predictions will be discovered.

\section{Discussion}

\noindent
Quantum mechanics was originally discovered as a description for 
atomic physics with no anticipation that it would be useful in theoretical
cosmology. The first such use was several decades later based on quantum
perturbations in the early universe seeding the growth of large scale
structure; this is well-known in inflationary scenarios and less known
in cyclic bounce models.

\bigskip

\noindent
In the present article we have suggested a second 
use of quantum mechanics in cosmology.
It arose by better understanding the CBE model.
The underlying quantum theory is based on AdS/CFT duality 
and entropic entropy where a source of encouragement is the emergent understanding 
of AdS/CFT and the Ryu-Takayanagi formula
from analysis of holographic quantum error-correcting codes.

\bigskip

\noindent
It is a decade since publication of BF which did generate a lot of discussion at that time
In addressing Tolman's no-go theorem, the present improved suggestion is to make
an infinitely-cyclic cosmology by requiring infinite cyclicity not of the
entropy of the whole universe but only of the quantum entanglement entropy
of the introverse, $S(int)$.

\bigskip

\noindent
This has the advantage that when $S(int)=0$
at turnaround, the introverse becomes a spacetime manifold
disconnected from the extroverse and hence the latter
can be jettisoned with impunity. This avoids the appearance
of an infiniverse after infinitely many cycles
and enables a model with only one universe.

\bigskip

\noindent
\section{Afterword}

\noindent
Just to add a few final words on a different topic which is the idea
that dark matter constituents are PIMBHs(=Primordial Intermediate Mass
Black Holes).  Our proposal \cite{FramptonDM} is that the Milky Way 
contains between ten million and ten billion massive black holes 
each with between twnety and a hundred thousand 
times the solar mass. Assuming the halo
is a sphere of radius a hundred thousand light years the typical separation
is between one hundred and one thousand light years which is also the most
probable distance of the nearest PIMBH to the Earth.

\bigskip

\noindent
Of the detection methods discussed in \cite{FramptonDM}, 
extended microlensing observations
seem the most promising and an experiment to detect
higher longevity microlensing events is being actively pursued.
The wide-field telescope must be in the Southern Hemisphere
to use the Magellanic Clouds (LMC and SMC) for sources.

\noindent
The best existing (since 1986) telescope is
Blanco 4m fitted with the DECam
having 520 Megapixels.
The future LSST(= Large Synoptic
Survey Telescope) under construction in Northern Chile
will not take first light until 2022.  It will be 8.4m with a 3200 Megapixel
camera. Long before LSST, at the Blanco 4m in Chile,
there is an approved NOAO proposal\cite{Kinney}
for a microlensing experiment which already
started February 2018 and expects results by
the end of 2019.

\subsection{The Reason for Dark Matter}

\noindent
The question why there is dark matter seems to us to be equally as important
as what is the dark matter, and there is more discussion in \cite{MagnumOpus}
The answer to why is the second law of thermodynamics applied to the entropy of the universe.
It is not a sharp argument since it depends on dynamics, and to make it rigorous would 
require solving Boltzmann's equations for every particle in the universe which is
impracticable. Nevertheless, it is strongly suggestive and we believe it can be the
correct reason.
The entropy from the SMBHs at galactic cores gives $S/k \sim 10^{103}$
and the identification $DM=PIMBHs$ can increase this to $S/k \sim 10^{106}$
depending on the PIMBH mass function and the dark matter was formed in the
first three years.

\end{document}